\documentclass[prb,showpacs,preprintnumbers,twocolumn]{revtex4}
\usepackage{tabularx}
\usepackage{amsmath}
\usepackage{amssymb}
\usepackage{graphicx}
\usepackage{dcolumn}
\usepackage{ulem}
\usepackage{bm}

\begin{document}

\title{Majorana Fermions Under Uniaxial Stress in
Semiconductor-Superconductor Heterostructures}
\author{Ming Gong$^{1,2}$}
\author{Li Mao$^{1,2}$}
\author{Sumanta Tewari$^{3}$}
\author{Chuanwei Zhang$^{1,2}$}
\thanks{Corresponding Author; Email: chuanwei.zhang@utdallas.edu}

\begin{abstract}
Spin-orbit coupled semiconductor nanowires with Zeeman splitting in
proximity contact with bulk $s$-wave superconductivity have recently been
proposed as a promising platform for realizing Majorana fermions. However,
in this setup the chemical potential of the nanowire is generally pinned by
the Fermi surface of the superconductor. This makes the tuning of the
chemical potential by external electrical gates, a crucial requirement for
unambiguous detection of Majorana fermions, very challenging in experiments.
Here we show that tunable topological superconducting regime supporting
Majorana fermions can be realized in semiconductor nanowires using uniaxial
stress. For n-type nanowires the uniaxial stress tunes the effective
chemical potential, while for p-type systems the effective pairing may also
be modified by stress, thus significantly enhancing the topological minigap.
We show that the required stress, of the order of 0.1\%, is within current
experimental reach using conventional piezo crystals.
\end{abstract}

\affiliation{$^{1}$Department of Physics, the University of Texas at Dallas, Richardson,
TX, 75080 USA \\
$^{2}$Department of Physics and Astronomy, Washington State University,
Pullman, WA, 99164 USA\\
$^{3}$Department of Physics and Astronomy, Clemson University, Clemson, SC,
29634 USA}
\maketitle

Majorana fermions (MFs) are quantum particles which are their own
antiparticles \cite{majorana37, Wilczek,Frantz}. MFs are not only of
fundamental interest because of their non-Abelian exchange statistics, but
also may serve as building blocks for fault-tolerant topological quantum
computation \cite{Kitaev2003,Nayak}. In the past few years, the possibility
of realizing MFs using quasiparticles in exotic solid state systems \cite%
{Read,Sarma,Bonderson,Stern,Ruthenate} has generated a lot of excitement in
the condensed matter community. In particular, it has been proposed recently
that MFs can be generated using a heterostructure consisting of two very
conventional materials: an $s$-wave superconductor and a semiconductor thin
film/nanowire with strong spin-orbit coupling \cite%
{J.Sau,J.Alicea,J.Sau3,Potter,Oreg,Roman,Mao1,Mao2,LiMao,Review-Beenakker}.
This proposal, following on the earlier idea proposed in cold atoms systems
\cite{Zhang}, has attracted widespread theoretical interest as well as
serious consideration in experiments. Very recently, following the
theoretical proposals, some preliminary experimental signatures \cite%
{Satoshi,Williams,Mourik,Deng,Rokh,Das} which may be related to the
existence of MFs have been observed, although the unambiguous detection of
MFs still remains an outstanding experimental challenge.

\begin{figure}[tbp]
\centering
\includegraphics[width=3.3in]{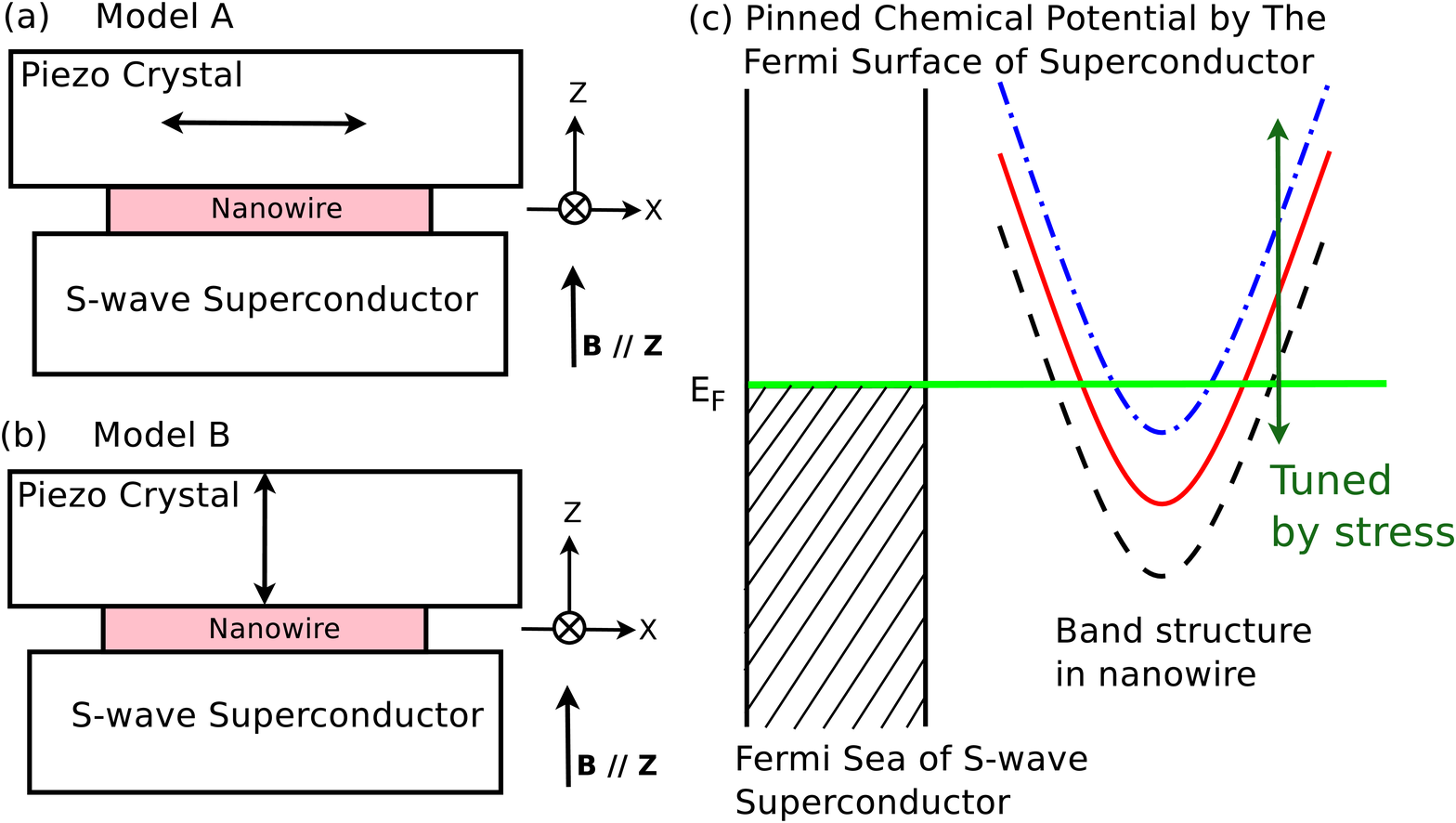}
\caption{(Color online). Engineering Majorana fermions via uniaxial stress.
The stretching direction is along $x$ in model A (a) and $z$ in model B (b).
The stress is assumed to be provided by conventional piezo crystals. (c)
shows the pinning of the chemical potential in semiconductor by the Fermi
surface of the $s$-wave superconductor. However, the total band structure of
the semiconductor can still be shifted by uniaxial stress.}
\label{fig-setup}
\end{figure}

In the proposed semiconductor heterostructures, MFs only exist for
topological superconducting states within a certain constrained parameter
regime \cite{J.Sau3}. Therefore, it is crucially important to have the
ability to tune various physical parameters in experiments. Since the
spin-orbit coupling strength and the size of the nanowire cannot be tuned
after the samples are fabricated, and the Zeeman field $V_{z}$ is generally
limited to a narrow window ($|V_{z}|\sim \Delta $, $\Delta $ is the
superconducting pairing) because of its possible depairing effect, it is
essential to be able to tune the chemical potential to the correct level for
the realization of the topological state. For semiconductor nanowires, the
chemical potential can be tuned by using suitably placed external electrical
gates. Unfortunately, the same technique may not work well in the proposed
Majorana system because the nanowire is in proximity contact with the
superconductor which has an extremely high carrier density. The pinning of
the chemical potential of the nanowire by the Fermi surface of the
superconductor thus poses a major challenge in experiments \cite{J.Sau2}. In
the recent Delft experiments \cite{Mourik} it has been found that very large
electric gate voltages ($\sim (10^{4}-10^{5})\Delta $) are required to tune
the topological quantum phase transition, which likely provides strong
evidence for the pinning of the nanowire chemical potential. The main
motivation of the present work is to provide an alternative method to
overcome this experimental difficulty which may greatly facilitate future
experimental searches of MFs in semiconductor-based heterostructures.

In this Rapid Communication we show that the topological superconducting
regime in both n- and p-type nanowires can be externally tuned using
uniaxial stress, which can be generated and controlled even by conventional
piezo crystals. The uniaxial stress can modify the band structure of the
nanowires slightly after the heterostructure system is fabricated,
remarkably leading to a topological transition from the trivial to the
topological superconducting state with MFs at the wire-ends. With the
experimentally accessible strength of the uniaxial stress, the effective
chemical potential can be tuned about 42 meV for electrons, and 5 - 20 meV
for hole levels. Moreover, for the p-type systems, the uniaxial stress may
also significantly enhance the minimum topological energy gap (minigap) that
protects the MFs from thermal excitations. The newly added elements for
generating the uniaxial stress can be effectively integrated into the design
of semiconductor devices using modern nanotechnology. Therefore, our
proposed scheme can go a long way in facilitating the realization and
detection of Majorana fermions in semiconductor quantum wire
heterostructures and the eventual implementation of topological quantum
computation.

Our basic setup for experiments is illustrated in Fig. \ref{fig-setup}a and %
\ref{fig-setup}b. The semiconductor nanowire (e.g., InSb, InAs, etc.) is in
proximity contact with an $s$-wave superconductor. The uniaxial stress
applied on the semiconductor nanowires can be generated using
nano-ferroelectric materials or by simply gluing the nanowire tightly to the
surface of piezoelectric crystals \cite{Shayegan, Thiele, Plumhof, Jons,
Seidl} such as the piezoelectric lead zirconic titanate (PZT) ceramic stack.
The stretching direction of the piezo crystal can be chosen either along $x$
(model A) or along $z$ direction (model B). The strain tensor can be
determined as,
\begin{eqnarray}
&&\varepsilon _{xx}^{(a)}=-\varepsilon ,\quad \varepsilon
_{yy}^{(a)}=\varepsilon _{zz}^{(a)}={\frac{2C_{12}}{C_{11}}}\varepsilon ,
\notag \\
&&\varepsilon _{zz}^{(b)}=-\varepsilon ,\quad \varepsilon
_{yy}^{(b)}=\varepsilon _{xx}^{(b)}={\frac{2C_{12}}{C_{11}}}\varepsilon ,
\label{eq-stress}
\end{eqnarray}%
where $\varepsilon =(1-a/a_{0})$ defines the relative changes of the lattice
constant along the corresponding crystallographic directions. Here $a_{0}$
and $a$ are the equilibrium and distorted lattice constants, respectively. $%
\varepsilon >0(<0)$ corresponds to compressive (tensile) stress. In
experiments the sign of $\varepsilon $ can be controlled by the voltage bias
across the piezo crystals \cite{Shayegan,Thiele,Plumhof,Jons,Seidl}. $C_{11}$
and $C_{12}$ are the elastic stiffness tensors. The superscripts in Eq.~(\ref%
{eq-stress}) represent the two different models shown in Fig. \ref{fig-setup}%
. Note that $\varepsilon \sim \mathcal{P}/\mathcal{Y}$, where $\mathcal{P}$
is the stress and $\mathcal{Y}$ is Young's modulus. Using typical values for
$\mathcal{Y}\sim 100$ GPa and $\mathcal{P}\sim 100$ MPa, we see that $%
\varepsilon \sim 0.1\%$. Such a small strain can be provided using
conventional piezo crystals. For model A, $\varepsilon $ up to $0.11\%$ has
already been realized in experiments, and in principle, $\varepsilon $ up to
0.6\% can be achieved \cite{Thiele,Plumhof}. For model B, it is more
suitable to provide compressive stress along the $z$ direction, and there is
no limitation on the maximum $\varepsilon $ because the compressive stress
is not limited by the gluing technique. We assume $|\varepsilon |<0.3\%$ the
most probable regime that can be accessed in experiments. Since the lattice
deformation is very small, the uniaxial stress has negligible effects on the
superconductor transition temperature as well as the s-wave pairing symmetry
\cite{Yang,Chen}, two properties that are crucial for the generation of
Majorana fermions.

%Following the original Delft experiments \cite{Mourik},
%here we use parameters for InSb nanowires\cite{Vurgaftman} to demonstrate
%that such a small strain is already sufficient for the realization of MFs.

\begin{figure}[tbp]
\centering
\includegraphics[width=3in]{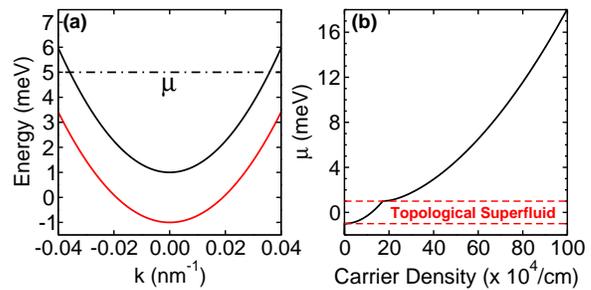}
\caption{(Color online) (a) Band structure of InSb nanowires. The solid
horizontal dash-dotted line represents the possible pinned chemical
potential when placed in proximity to the superconductor. (b) The chemical
potential as a function of carrier density. $\protect\alpha \sim 0.2$ eV$%
\cdot $\AA , $m=0.013m_{0}$, $\Delta _{0}=0.5$ meV, $V_{z}=1.0$ meV.
Parameters are from Ref. \onlinecite{Vurgaftman}.}
\label{fig-electron}
\end{figure}

The effective Hamiltonian for n-type nanowire under uniaxial stress reads as
\cite{Wei, Gong},
\begin{equation}
H=[{\frac{k^{2}}{2m}}+\alpha (\mathbf{p}\times \mathbf{\sigma }%
)_{z}+V_{z}\sigma _{z}+a_{c}\text{Tr}(\mathbf{\varepsilon })]-\mu _{F}.
\label{eq-electron}
\end{equation}%
where $a_{c}$ denotes the deformation potential of the conduction band, $%
\alpha $ is the spin-orbit coupling strength, and $V_{z}$ is the external
Zeeman field induced by the magnetic field. $\mu _{F}$ is the true chemical
potential of the semiconductor that is pinned to the Fermi surface of the
superconductor. The uniaxial stress does not change the band structure, but
shifts the effective chemical potential to $\mu =\mu _{F}-a_{c}\text{Tr}(%
\mathbf{\varepsilon })$, through which the topological region $%
V_{z}^{2}>\Delta ^{2}+\mu ^{2}$ \cite{Oreg,Roman} may be achieved. In Fig. %
\ref{fig-electron}a, we plot the typical band structure of free electrons in
nanowires in a single transverse confinement band and in Fig. \ref%
{fig-electron}b we plot the corresponding chemical potential as a function
of carrier density. The topological superconductivity with MFs can be
achieved when the chemical potential falls in the small window in Fig. \ref%
{fig-electron}b, in which case the system cuts only one Fermi surface. The
small window in the parameter space greatly limits the flexibility for the
experimental observation of MFs. For InSb, $a_{c}\sim -6.94$ eV, and $%
|\varepsilon |<0.3\%$, we estimate $a_{c}\text{Tr}(\varepsilon )\sim \pm 21$
meV, the same magnitude as the energy difference between the chemical
potential of the nanowire and the conduction band minima in the Delft
experiment \cite{Mourik}. We see from Fig. \ref{fig-electron}b that such a
huge change of the effective chemical potential can change the carrier
density by about 1 - 2 orders of magnitude. Thus, for a wide range of
carrier density, the nanowire can always be tuned to the topological regime
in the experiment.

\begin{figure}[tbp]
\centering
\includegraphics[width=3.3in]{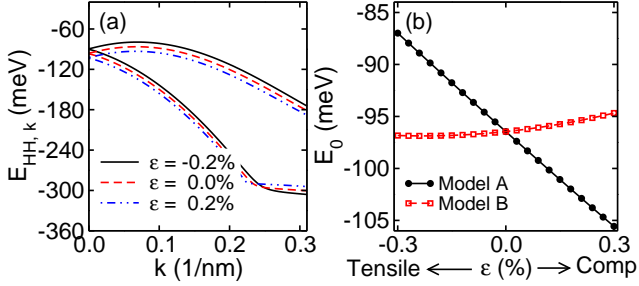}
\caption{(Color online). (a) Typical band dispersions of the heavy hole
bands for different values of strain $\protect\varepsilon $ in model A. (b)
Energy of heavy hole bands at $k=0$ as a function of strain $\protect%
\varepsilon $ for models A and B. Other parameters are $\protect\gamma %
_{1}=34.8$, $\protect\gamma _{2}=15.5$, $\protect\alpha =2.0$ eV$\cdot $\AA %
, $V_{z}=1.5$ meV, $L_{z}=14$ nm, and $L_{y}=10$ nm, $\protect\mu =0$.
Parameters are from Ref. \onlinecite{Vurgaftman}.}
\label{fig-bands}
\end{figure}

The Hamiltonian for p-type nanowires (assuming the sizes along the $y$ and $%
z $ directions are $L_{y}$ and $L_{z}$, respectively) under uniaxial stress
reads as \cite{Bernevig},
\begin{equation}
H=H_{\text{KL}}+H_{\text{BP}}-\mu _{F},
\end{equation}%
where $H_{\text{KL}}={\frac{(2\gamma _{1}+5\gamma _{2})}{4}}\nabla
^{2}-\gamma _{2}(\nabla \cdot \mathbf{J})^{2}-i\alpha (\mathbf{J}\times
\nabla )_{z}+V_{z}J_{z}$ is the Kohn-Luttinger Hamiltonian, $\mathbf{J}$ is
the total angular momentum operator for the spin-3/2 holes and $\gamma _{1}$
and $\gamma _{2}$ are Luttinger parameters. The second term describes the
Bir-Pikus model \cite{Wei,Gong,He,Pryor}
\begin{equation}
H_{\text{BP}}=%
\begin{pmatrix}
P_{\varepsilon }+Q_{\varepsilon } & 0 & R_{\varepsilon } & 0 \\
0 & P_{\varepsilon }-Q_{\varepsilon } & 0 & R_{\varepsilon } \\
R_{\varepsilon }^{\ast } & 0 & P_{\varepsilon }-Q_{\varepsilon } & 0 \\
0 & R_{\varepsilon }^{\ast } & 0 & P_{\varepsilon }+Q_{\varepsilon }%
\end{pmatrix}%
,
\end{equation}%
where $P_{\varepsilon }=-a_{v}\text{Tr}(\varepsilon )$, $Q_{\varepsilon }=-{%
\frac{b}{2}}(\varepsilon _{xx}+\varepsilon _{yy}-2\varepsilon _{zz})$, $%
R_{\varepsilon }={\frac{\sqrt{3}}{2}}b(\varepsilon _{xx}-\varepsilon _{yy})$%
, with $a_{v}$ and $b$ the deformation potentials of the valence bands.
Notice that $P_{\varepsilon }$, $Q_{\varepsilon }$ and $R_{\varepsilon }$
have totally different roles to the band structures of holes in nanowires. $%
P_{\varepsilon }$ shifts the global band structure, while $Q_{\varepsilon }$
increases or decreases the splitting between the heavy hole (HH) and light
hole (LH) bands. The non-zero $R_{\varepsilon }$ may greatly enhance or
suppress the coupling between HH and LH, thus modifying the effective
pairing strength. In contrast to the case of electrons, the two models A and
B yield totally different results.

The modification of the band structure of p-type nanowires due to uniaxial
stress is shown in Fig. 3a for different values of $\varepsilon $. Here we
only plot the two HH bands because the LH bands are separated by a large
energy gap ($\sim $ 100 meV) induced by the confinement. At $k=0$, two HH
bands are split by a small Zeeman field $V_{Z}$. When the chemical potential
lies in the Zeeman gap, the system has only a single Fermi surface, yielding
topological superconductivity and the associated MFs under suitable
conditions. We see from Fig. 3 that by tuning the uniaxial stress, we can
shift the bands of the semiconductor up or down so that the Fermi level of
the superconductor can lie in the Zeeman gap, yielding topological
superconductivity. In Fig. 3b, we plot $E_{0}$, the energy of the HH bands
at $k=0$, against $\varepsilon $. Within the experimentally accessible
regime, $E_{0}$ can be tuned by $\pm 3$ meV for model B and $\pm 10$ meV for
model A.

To obtain concrete parameter regions of the MFs, the superconducting order
parameter need be taken into account. Because the uniaxial stress only
shifts the effective chemical potential for the n-type of semiconductor, we
consider only p-type nanowires. In the nanowire heterostructures, the
superconducting order parameter can be induced to the nanowire through
proximity effect \cite{Kouwenhoven1,Xiang}, yielding $H_{\text{sc}}=\sum_{m={%
\frac{1}{2}},{\frac{3}{2}}}\int d\mathbf{r}\Delta _{m}\psi _{m}^{\dagger
}\psi _{-m}^{\dagger }$. The corresponding BdG Hamiltonian can be written as
\cite{LiMao}
\begin{equation}
H_{\text{BdG}}=%
\begin{pmatrix}
H_{\text{1D}} & \Delta _{4} \\
\Delta _{4}^{\ast } & -\mathcal{\gamma }^{\dagger }H_{\text{1D}}^{\ast }%
\mathcal{\gamma }%
\end{pmatrix}
\label{eq-BdG}
\end{equation}%
in the Nambu spinor basis $\Psi =(\psi ,\mathcal{\gamma }\psi ^{\dagger
})^{T}$. Here $H_{\text{1D}}=\int dydz\psi _{y}^{\ast }\psi _{z}^{z}H\psi
_{z}\psi _{y}$, $\mathcal{\gamma }=i(I\otimes \sigma _{x})\tau _{y}$, $%
\sigma _{x}$, $I$ and $\tau _{y}$ are Pauli operators, and $\Delta _{4}=$%
diag($\Delta _{\frac{3}{2}}$, $\Delta _{\frac{1}{2}}$, $\Delta _{\frac{1}{2}%
} $, $\Delta _{\frac{3}{2}}$).

The topological parameter regime for MFs can be obtained by the topological
index \cite{Kitaev,Ghosh}
\begin{equation}
\mathcal{M}=\text{sign}(\text{Pf}(\Gamma (0))\cdot \text{Pf}(\Gamma ({\frac{%
\pi }{a}})))
\end{equation}%
where Pf$(\Gamma )$ refers to the Pfaffian of the matrix $\Gamma =-iH_{\text{%
BdG}}(k)(\tau _{y}\otimes \mathcal{\gamma })$, $a$ is the lattice constant. $%
\mathcal{M}=+1$ (-1) corresponds to the topologically trivial (nontrivial)
superconducting states without (with) MFs. Note that for sufficient large $k$
all the eigenvalues of $\Gamma $ are dominated by the $k^{2}$ terms,
yielding s$\text{ign}(\text{Pf}(\Gamma ({\frac{\pi }{a}})))=1$. The Pfaffian
at $k=0$ can be derived analytically, yielding $\mathcal{M}=\text{sgn}%
\mathcal{F}$, with $\mathcal{F}=f_{0}-f_{1}V_{z}^{2}+{\frac{9}{16}}V_{z}^{4}$%
, $f_{0}=(\bar{\mu}^{2}+\Delta _{{\frac{3}{2}}}\Delta _{{\frac{1}{2}}}-\beta
_{1}^{2}-\beta _{2}^{2})^{2}+((\Delta _{{\frac{3}{2}}}-\Delta _{{\frac{1}{2}}%
})\bar{\mu}+\beta _{1}(\Delta _{{\frac{3}{2}}}+\Delta _{{\frac{1}{2}}}))^{2}$%
, $f_{1}=(10\bar{\mu}^{2}+10\beta _{1}^{2}+16\beta _{1}\bar{\mu}+9\Delta _{{%
\frac{1}{2}}}^{2}+\Delta _{{\frac{3}{2}}}^{2}-6\beta _{2}^{2})/4$, $\beta
_{1}=\pi ^{2}\gamma _{2}(L_{z}^{-2}-L_{y}^{-2}/2)+Q_{\varepsilon }$, $\beta
_{2}=\sqrt{3}\pi ^{2}\gamma _{2}L_{y}^{-2}/2+R_{\varepsilon }$, and $\bar{\mu%
}=\mu +\gamma _{1}\pi ^{2}(L_{y}^{-2}+L_{z}^{-2}/2)$ \cite{SI}. The boundary
for topological phase transition is determined by $\mathcal{F}=0$.
Generally, the magnitudes of $\Delta _{{\frac{3}{2}}}$ and $\Delta _{{\frac{1%
}{2}}}$ are not essential for the topological quantum phase transition (but
the relative sign is important) \cite{LiMao}. Henceforth we only consider
two different possible cases (I) $\Delta _{{\frac{3}{2}}}=\Delta _{{\frac{1}{%
2}}}=\Delta $ and (II) $\Delta _{{\frac{3}{2}}}=-\Delta _{{\frac{1}{2}}%
}=\Delta $. For other values of $\Delta _{{\frac{3}{2}}}$ and $\Delta _{{%
\frac{1}{2}}}$, the results are similar.

\begin{figure}[tbp]
\centering
\includegraphics[width=3.3in]{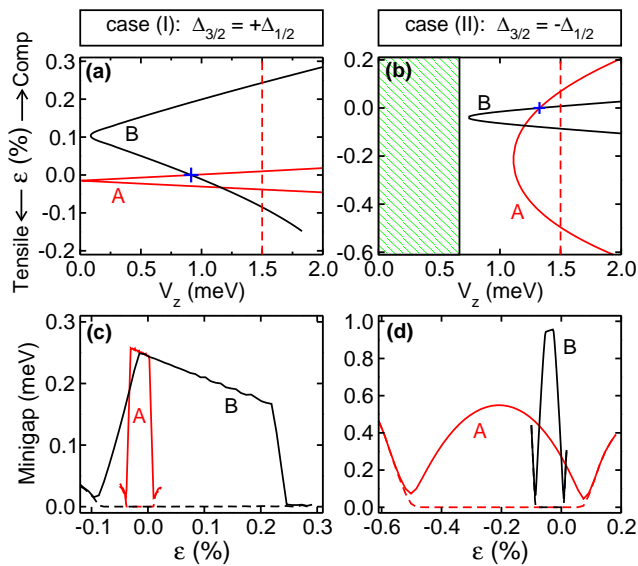}
\caption{(Color online). The parameter regimes for the existence of Majorana
fermions. (a), (b) show the results for case (I) and case (II),
respectively, see text for details. The plus sign ($+$) corresponds to $%
V_{z} $ without uniaxial stress. For case (I) we use $L_{y}\simeq 10.0$ nm, $%
L_{z}\simeq 14.0$ nm, for case (II) we use $L_{y}\simeq 14.2$ nm, $%
L_{z}\simeq 9.7$ nm. The chemical potential in all the figures are pinned at
$\protect\mu =-96.0$ meV. In (b) the shaded regime corresponds to the lower
bound of $V_{z}/\Delta =2/3$. The corresponding mini-gap and lowest
non-negative energy level obtained by numerically solving the BdG Eq. (%
\protect\ref{eq-BdG}) are presented in (c) and (d), respectively, for $%
V_{z}=1.5$ meV. The other parameters are the same as Fig. \protect\ref%
{fig-bands}.}
\label{fig-phase}
\end{figure}

In Fig. \ref{fig-phase}a and \ref{fig-phase}b, we plot the boundary between
the topological and non-topological superconducting states. In these
figures, we assume that without uniaxial stress the chemical potential lies
in a regime which requires a large Zeeman field for realizing the
topological superconducting state. By applying the uniaxial stress the
required critical Zeeman field can be significantly reduced. In case (I) in
Fig. \ref{fig-phase}a, the critical Zeeman field can even approaches zero
for model A. We have also verified that for a wide range of parameters ($\mu
$, $L_{z}$, $L_{y}$ $\cdots $) similar features can always be found. For
case (II) in Fig. \ref{fig-phase}b the required Zeeman field can be reduced
to around 1 meV ($B_{z}=0.3$ T for $g_{h}^{\ast }\sim 50$). Generally for
case (II), the minimum required $V_{z}^{c}\simeq p\Delta $, where $p=2\sqrt{%
1+\beta _{1}^{2}/\beta _{2}^{2}}/(2+\sqrt{1+\beta _{1}^{2}/\beta _{2}^{2}}%
)\in \lbrack 2/3,2)$. We see that there are also a wide range of parameters
that enable us to achieve the minimum required Zeeman field at $p=2/3$ via
uniaxial stress. To further verify that the right regime of each curve in
Fig. \ref{fig-phase}a and \ref{fig-phase}b are indeed the topological
superconducting regime we plot the mini-gap (solid line) and the lowest
non-negative energy level (dashed line) as a function of uniaxial stress $%
\varepsilon $ in Figs. \ref{fig-phase}c and \ref{fig-phase}d, respectively.
In the topological superconducting state, the zero energy MFs indeed exist
with large minigaps around several Kelvin. We have also confirmed that the
corresponding wavefunctions of MFs are well localized at the two ends of the
nanowire.

The difference between the two types of uniaxial stress in models A and B
can be understood by projecting the Hamiltonian to the lowest two HH bands
\cite{SI,winkler,Bernevig,Wei,Gong,He,Pryor}, which yields the effective
pairing at $k\rightarrow 0$,
\begin{equation}
\Delta _{\text{eff}}=(\Delta _{{\frac{3}{2}}}-\kappa \Delta _{{\frac{1}{2}}%
})/(\kappa +1)
\end{equation}%
where $\kappa =(\sqrt{(\beta _{1}/\beta _{2})^{2}+1}-\beta _{1}/\beta
_{2})^{2}\in (0,\infty )$. When $\varepsilon =0$, $\kappa $ only depends on
the size of the nanowire, thus cannot be tuned. However, when the uniaxial
stress is applied, $\kappa $ can be tuned in a considerably wide range. For
model A, the off-diagonal term $R_{\varepsilon }\neq 0$, thus $\beta _{2}$
can approach zero with a properly chosen strain $\varepsilon $. For the
parameters used in Fig. \ref{fig-phase} we find that the effective pairing
increases (decreases) monotonically as a function of $\varepsilon $ for
model A (B), thus for model B, we observe significant enhancement of the
mini-gap ($\sim $ 30\%) in Fig. \ref{fig-phase}d. The maximum increase of
the mini-gap can be obtained by optimizing different physical parameters.

Finally, several remarks are in order. First, the same idea discussed above
for a single band model can be straightforwardly extended to the multiband
case. Using the diameter of nanowire from Ref. \cite{Mourik}, we estimate
the band spacing for electron (hole) is $\sim 6$ ($\sim 3$) meV, with $\sim
4 $ bands occupied. Thus we expect that the stress can tune the effective
chemical potential of both electrons and holes to the topological regime
even though the initial value corresponds to an even number of Fermi
surfaces. Second, our proposal here can also be used to engineer MFs in the
vortex core of semiconductor quantum wells. For electrons the tuning of the
band structures is exactly the same as in Eq. \ref{eq-electron}. For holes
there are some qualitative difference from Fig. 3 since the confinement
along the $y$ direction is relaxed. As a consequence, $\beta _{2}=0$ when $%
\varepsilon =0$, thus $R_{\varepsilon }$ play a more significant role in the
determination of the minigap of MFs. Third, we have also checked the
validity of our proposal for InAs nanowires, and similar features have been
found. However, for InAs, we note that\cite{Vurgaftman} $a_{c}\sim -5.17$
and $b\sim -1.00$, which are smaller than their counterparts in InSb, thus a
slightly larger stress is required.

To conclude, due to the proximity effect between nanowires and a bulk
superconductor, the chemical potential of the nanowire is generally pinned
by the Fermi surface of the superconductor. Consequently, tuning the
chemical potential of nanowires via electrical gates to bring it in the
topological regime is inefficient in this setup. We show that this crucial
obstacle can be overcome using experimentally accessible uniaxial stress
which modifies the band structure slightly, leading, remarkably, to a
transition from non-topological to topological states with MFs. The newly
added elements for generating uniaxial stress can be effectively integrated
into the design of semiconductor devices using modern nanotechnology.
Therefore our scheme can be used for the realization of topological Majorana
fermion excitations in semiconductors and the implementation of topological
quantum computation.

\textbf{Acknowledgement:} We thanks valuable discussions with K. D. Jons and
J. D. Plumhof about the experimental implementation of unaxial stress and P.
Kouwenhoven about the possible pinning of chemical potential in their
experiments. This work is supported by DARPA-MTO (FA9550-10-1-0497),
DARPA-YFA (N66001-10-1-4025), and NSF-PHY (1104546 and 1104527).

\end{document}